# Estimating the EMG response exclusively to fatigue during sustained static maximum voluntary contraction

Jing Chang[1,2], Damien Chablat[1], Fouad Bennis[1], Liang Ma[2]

[1] IRCCyN, Ecole Centrale de Nantes, 44321 CEDEX 3, Nantes, France
[2] Department of Industrial Engineering, Tsinghua University, Beijing, 100084,P.R.China
{Jing.chang, Damien.Chablat, Fouad.Bennis} @irccyn.ec-nantes.fr;
liangma@tsinghua.edu.cn

**Abstract.** The increase of surface electromyography (sEMG) root-mean-square (RMS) is very frequently used to determine fatigue. However, as RMS is also influenced by muscle force,its effective usage as indicator of fatigue is mainly limited to isometric, constant force tasks.This research develops a simple methodto preclude the effect of muscle force, hereby estimates the EMG amplitude response exclusively to fatigue with RMS. Experiment was carried out on the biceps brachiis of 15 subjects (7males, 8 females) during sustained static maximum voluntary contractions (sMVC).Result shows that the sEMG RMS response to fatigue increasesto 21.27% while muscle force decreasing to 50%MVC, which implies that more and more extra effort is needed as muscle fatigue intensifies. It would be promising to use the RMS response exclusively to fatigue as an indicator of muscle fatigue.

## 1 Introduction

Muscle fatigue is a complicate phenomenon believed to be closely related to musculoskeletal disorder[1]. In ergonomics, fatigue is usually measured from three aspects:the reduction of force output capability, the self-rating discomfort, as well as the myoelectric activity change.The reduction of force output capabilityhas been well modeled[2], and been used to describe fatigueand calculate fatigue resistance[3,4]; the self-rating discomfort was reported to be a valid estimator[5], but its usage might be limited due to the subjectivity. The myoelectric activitymonitored by surface EMG (sEMG) shows the in situ and real-time changes of the muscles[6]. Therefore, ithas been frequently used to determine fatigue[7].

  Overall, sEMG signal is the synthesis of myoelectric and anatomic properties: the shape of motor unit action potential (MUAP) waveform[8], the firing rate of MUAP trains, the recruitment of new motor units[9]as well as thesubcutaneous tissues which act as a volume conductor and cause a spatial low-pass filtering effect when conducting myoelectric signals[10].In the process of fatigue, metabolites such as lactic acid concentrate, for which the muscle fibre conduction velocity (CV) decreases. The decrease of CV directly changes the shape of the MUAP waveform and leads to the compression of EMGspectrumtowards its lower part[11]. As a result, more energy passes through the tissues which act as a low pass filter, and is detected by the surface electrodes, leading to an increase of sEMG amplitude.

The changes ofsEMG characteristic values,such as decrease of the median power frequency (MDF) and increase of the root-mean-square (RMS), are very often used to estimate fatigue[12,13,14]. But it should be noted that thesesEMGindicators are not always applicable. Firstly,sEMG spectral compressions are sometimes slight and insignificant statistically, especially during low force level tasks[5,15].As De Luca[7] puts it, the usage of spectral variables should be limited toisometric, constant-load tasks greater than 30% of subjects' maximal force. In addition, thesEMG amplitude is significantly affected bymuscle force: greater force leads to larger amplitude[16,17]. It seemsthe RMS should only be used in constant-force fatigue process.

However, the tasks that are cared most aredynamic, inconstant-force ones. These tasks are characterized by time-varying changes in forces exerted variations as well as in working postures[12]. Both of the two factors lead to EMG changes despite of fatigue. To indicate fatigue with sEMG in real operations, it is logistic to start from tasks with fixed posture and inconstant force, such as sustained static maximal exertions.Voluntary sustained static maximal exertions, or sustained maximum voluntary contractions (sMVC), are characterized by rapid muscle fatigue and continuous force decline[18]. In this process, the impact of intensifying fatigue on sEMG amplitude is overwhelmed by the impact of decreasing muscle force. As a result, the RMS turns to decline along with fatiguing progress[19].

In this research, a simple methodto preclude the impact of muscle force on RMS is tested during sMVC and the sEMG RMS response exclusively to fatigue is identified. The possibility of using the identified value as an indicator of fatigue is discussed.

## 2 Methods

### 2.1 Subjects

A number of 15 subjects (7 males, 8 females) took part in the experiment. They were healthy, aged between 20 and 35 and free from any upper limb pain during the previous 12 months. All of them are right-handed. Other criteria include moderate (non-extreme) level of self-reported daily physical activity. After being fully informed of the nature of the experiment, they signed an informed consent. Anthropometry data were measured upon their arrival at the laboratory (see Table 1).

**Table 1.**Subject physical characteristics.

| Characteristic | Mean | SD |
| --- | --- | --- |
| Age (year) | 28.3 | 4.7 |
| Height (cm) | 167.1 | 9.1 |
| Weight (kg) | 67.1 | 14.0 |
| BMI | 23.8 | 3.5 |
| Maximum Voluntary Contraction (N) | 137.8 | 47.8 |

## 2.2 Experiment setup

The experiment platform consisted of a dynamometer (BET®Primus RS), a customized chair without armrest, an electromyography recorder (TeleMyo 2400T V2®) and an interactive monitor. The muscle force and the surface EMG were recorded by the dynamometer and EMG recorder at frequency of 20 Hz and 1500 Hz respectively. Real-time muscle force was recorded by and displayed on the monitor.

In the experiment, each subject was told to be seated on the chair with torso and upper arms perpendicular to the ground. The elbow joint angle of the dominant arm was kept at 90 degree with forearm supine and horizontal. Subject's output strength was transduced to the dynamometer by a lever that is hold in hand (see Figure 1). Body movement was mechanically restrained by belts restraining legs, trunk and shoulder to the chair. During the whole experiment, subjects were closely monitored to maintain the posture as still as possible. In all the experiments, bar arms were used and the room temperature was maintained around 22°C by air-conditioning if necessary.

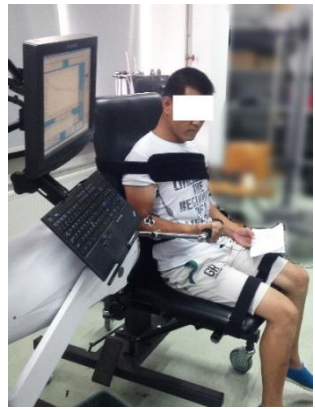

**Fig. 1.** Experiment posture.

## 2.3 Experiment protocol

Before experiment, subjects were trained to sustain submaximal muscle force without joint movements.
The experiment includes three sessions. First there was the initial Maximum Voluntary Contraction (iMVC) session. Subject's iMVC was obtained in the position described above. At least three short-time MVC trials (each lasting for 3s) were performed, with 10 min rest in between. The iMVC trials were continued until there were three measures whose Coefficient of Variation (CV) was less than 5%. The average value of the three measures was taken as iMVC.
After a recovery break of 10 minutes, there came the simple contraction sessions, where subjects perform a series of five short-time submaximal exertions, from 50% MVC to 90%MVC, in steps of 10% MVC. The sequence of the five exertions was generated by computer randomly for every subject. Each exertion lasted for only 3s to avoid fatigue. After each exertion, at least 5 minutes rest was taken until a complete recovery was reported by the subject. In this session, the interactive monitor that displayed the real-time force was set in front of the subjects and the target force level was clearly labelled on the screen.

Finally it was the sMVC fatigue session. After a total recovery of at least 15min, the subject was asked to exert his maximum strength to lift the transducing lever handle of the dynamometer and to sustain the maximum effort for 60 seconds. The force decline was recorded by the dynamometer automatically. Subjects received non-threatening verbal encouragement throughout the procedure.

### 2.4 Data collection

Surface EMG data was collected from the biceps brachii muscles by a pair of disposable Ag-AgCl electrodes. The electrodes were 1cm in diameter each, placed on the belly of the biceps brachii, with a 2.0 cm space in between. The skin was carefully shaved, cleared by alcohol and slightly abraded. Intra-electrodes resistance was kept below 10KΩ. A ground electrode was placed over the end of the humerusin the elbow flexor joint.

EMG signals were recorded continuously throughout the three test sessions. Raw signals were sampled at 1500 Hz. RMS was calculated using MATLAB.

### 2.5 Data analysis

For each 3s exertion of the simple contraction session, the RMS was calculated from the 1s-long EMG signals fragment in the central of the 3s-long signals (from 1s to 2s), noted as $RMS_s$.

During the sMVC, the muscle force declines from almost 100% MVC all the way along. For each subject, the time points when the force reaches 90%MVC, 80% MVC, 70% MVC, 60% MVC and 50% MVC were determined and the nearby EMG signals (1s-long) corresponding to these time points were selected. RMSs were calculated from these EMG signals fragments, noted as $RMS_f$.

By comparing the $RMS_f$ with $RMS_s$, one may figure out the EMG amplitude changes caused by fatigue, precluding the effect of muscle force. For each subject, all RMS values were normalized by the RMS of EMG signals when he or she reached the peak force during the sMVC fatigue session.

$$RMS_{fe} = RMS_f - RMS_s \qquad (1)$$

## 3 Results

### 3.1 Simple contraction session

Monotonous and significantly linear relationship (Pearson's test: $r = 0.74$; $p < 0.001$) was found between muscle force and RMS values in the simple contraction session

(shown in Figure 2). Larger muscle force corresponds stronger EMG signals.

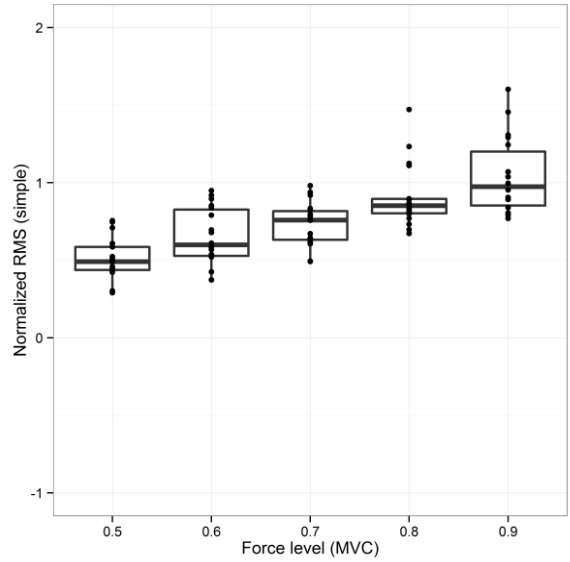

**Fig. 2.** Subjects' normalized RMS values during simple contraction session.

### 3.2 Fatigue sessions

Under the influence of both muscle force declines and the fatigue increases, the amplitude of EMG decreases along the fatigue process. As shown in Figure 3, from 90% MVC to 50% MVC, the $RMS_f$ goes from 95.4% to 70.2% (normalized by RMS corresponding to maximal force).

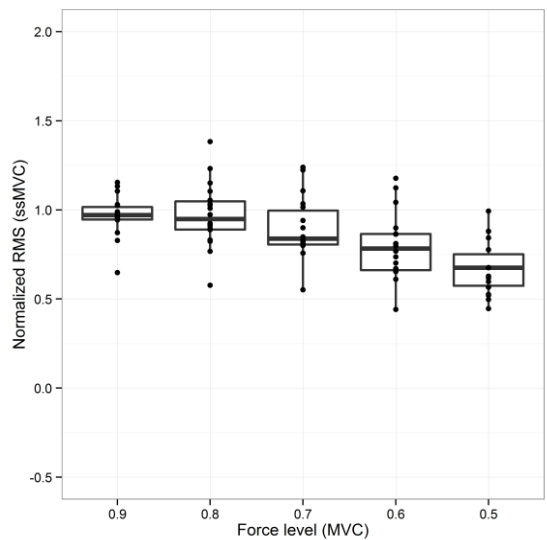

**Fig. 3.** Subjects' normalized RMS values during sMVC fatigue session.

**Table 2.** Normalized RMS values corresponding to five force levels during sMVC session

|  | 90% MVC | 80% MVC | 70% MVC | 60% MVC | 50% MVC |
|---|---|---|---|---|---|
| $RMS_f$*(%) | 95.38(13.55) | 96.37(19.58) | 89.69(18.30) | 80.06(14.37) | 70.19(16.90) |
| $RMS_{fe}$*(%) | 0.54(11.78) | 2.24 (19.18) | 14.70 (16.42) | 14.67 (18.88) | 21.27(18.66) |

* Normalized by the RMS values corresponding to maximal force.

### 3.3 RMS response exclusively to fatigue

After precluding the influence of muscle force by equation (1), sEMG RMS response exclusively to fatigue was determined. Result shows that it increases with fatigue strengthening: $RMS_{fe}$ goes from 0.54% up to 21.27% while muscle force reduces from 90% MVC to 50% MVC. Boxplot is shown in Figure 4.

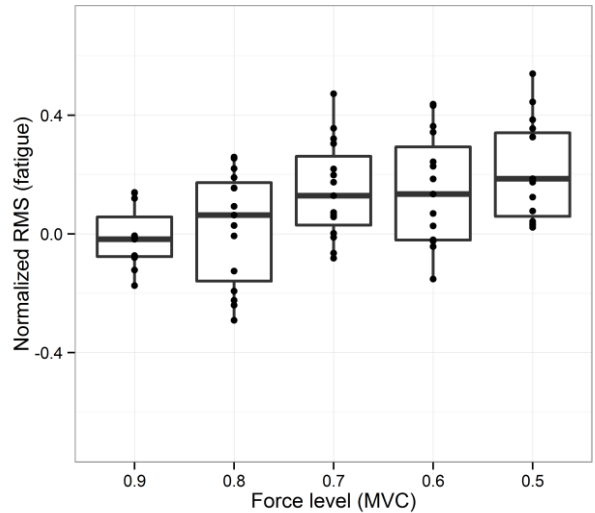

**Fig. 4.** Subjects' RMS response to fatigue during sMVC. Impact of muscle force is precluded.

## 4 Discussion

### 4.1 Force - RMS relationship in simple contraction

In this experiment, linear relationship was found between muscle forces and RMS values during the simple contraction session. This clear relationship between mechanical and the electrical responses of human muscle is well documented in previous researches[20,21,22] under voluntary isometric contractions.

As mentioned above, the RMS value of the sEMG signals is decided by: (1). the percentage of MUAP trains wave that filtered by the tissue and reach the electrodes,

$\tau_0$

designated as $p$; (2). the average length of the MUAP trains, designated as $\bar{l}$ ;(3). the firing rate of the MUAP trains, designated as $f$; (4). the meanamplitude of filtered MUAP trains, designated as $\bar{A}$. Then RMS could be indicated by equation (2) ( is the Constant coefficient).

$$RMS = \tau_0 \ \bar{l} \ p \ f \ \bar{A} \qquad (2)$$

During the simple contractions when the muscles are fresh, the shape of MUAP waveform and the filtering threshold of the tissue remain unchanged. Therefore $p$ is constant. With the increase of muscle force, MUs with higher firing rate of their amplitude potential trains are recruited and the firing rates of initial MUAP trains increase. These factors lead to the increases of $\bar{A}$ and $f$, and furthermore an increase of total RMS.

### 4.2 RMS during Fatigue Process

During the sMVC fatigue progress, the RMS reduces toabout 70% when muscle force reaches 50% MVC. Similar results have been reported in previous researches, as listed in Table 3. Generally, EMG amplitude reduced by30% to 70% during a maximal voluntary contraction sustained for 60 s, depending on different muscles and protocols.

**Table 3.** EMG amplitude changes during sustainedmaximal voluntary contraction.

| Authors | Muscle | Amplitude value | Tendency | Magnitude |
| --- | --- | --- | --- | --- |
| Stephens and Taylor[23] | First dorsal interosseous | srEMG* | Decrease | To 53% ± 7% |
| Bigland-Ritchie, et al.[19] | Adductor pollicis | srEMG* | Decrease | By50% to 70% |
| Kent-Braun[17] | Ankle dorsiflexor | iEMG** | Decrease | To 72.6% ± 9.1% |
| Bigland-Ritchie[24] | Adductor pollicis | srEMG* | Decrease | About 50% |

*srEMG - Smoothed rectified EMG;
**iEMG –Integreted EMG

As mentioned above, theRMSof sEMG has been found to increase in the process of constant-force fatiguing tasks[25,12,13]. When the body tries to maintain the target forces, a progressive increase of MUAP trains firing rate take place and MUs with larger amplitude are recruited[27], which increase the $f$ and $\bar{A}$ in Eq.2. At the same time, the percentage of filtered MUAP trains $p$ grows due to fatigue. As a result, the total RMS increases.

Whereas in the procedure of sustained maximal exertions, both acute muscle fatigue and rapid force decline are identified. The MUAP trains are pushed to change in two directions: on the one hand, in every moment of sMVC, the muscle is trying hard to maintain its original force, which leads to increases of $f$ and $\bar{A}$; on the other hand, as the muscle fails to maintain its original force, muscle force continue to decline, which brings about the decreases of $f$ and $\bar{A}$. As a consequence, the impact of force decline prevails over that of fatigue and decreases of $f$ and $\bar{A}$ are observed[27]. Despite of the increase of the percentage of filtered MUAP trains $p$, the total RMS declines with time.

### 4.3 EMG response exclusively to fatigue

After precludingthe impact of muscle force changes,the sEMG RMS response to fatigue is found to increase from 0.54%to 21.27%, along with muscle force decreasing from 90%MVC to 50%MVC.The underlying implication is that in this process, although the firing rate of and average amplitude of MUAP trains decline significantly, increasingly extra effort is made compared with that when the fresh muscle exerts same forces. The more severe the muscle fatigue is, the more extra effort it is needed.

It is notable from Table 2 that no extra RMS increase is detected when force declines from 70%MVC to 60%MVC. This should be explained by the different percentage held by MU deactivation and firing rate slowing in force reduction at different force output level. As the muscle force increases, MUs are recruited in the order of their firing rate and twitch tension from low to high[9,28]along with the average firing rate speeding up[29]. The recruitment of new MU was reported to terminate by 60%MVC[30] or 75%MVC[28] in biceps brachii. Conversely, during sMVC session, before the muscle force reaches about 70%MVC, the slowing of MU firing rate contributes the largest partto force loss. From 70%MVC to 60%MVC, the deactivation of MU takes part in.MUs with the largest twitch tension are deactivate, which makes a great contribution to force loss. Hereby, the firing rate slowing and RMS changes are not significant.

## 5 Conclusions

In this research, a simple way to identify the sEMG RMS response to fatigue was tested on biceps brachii muscles during sMVC process. The impact of muscle force changes on RMS is precluded by simply subtracting the RMS of the fresh muscle when exerting corresponding force from the total RMS.Result shows that the sEMG RMS response to fatigue increasesalong with the fatigue process to as much as 20%, which implies that more and more extra effort is needed as muscle fatigueintensifies. It would be promising to use the RMS response exclusively to fatigue as an indicator of muscle fatigue.

**Acknowledgments.** This work was supported by the National Natural Science Foundation of China under Grant numbers 71101079 and 71471095, by the Tsinghua University Initiative Scientific Research Program under Grant number 20131089234, by Chinese State Scholarship Fund, and by INTERWEAVE Project (Erasmus Mundus Partnership Asia-Europe) under Grants number IW14AC0456 and IW14AC0148.